# Reduction of the COSMOS Southern Sky galaxy survey data to the RC3 standard system

J. Rousseau[1], H. Di Nella[1,2], G. Paturel[1], C. Petit[1]

[1] CRAL-Observatoire de Lyon, F69561 Saint-Genis Laval, FRANCE,
[2] School of Physics, University of New South Wales, Sydney, NSW 2052, AUSTRALIA

**ABSTRACT**
After having cross-identified a subsample of LEDA galaxies in the COSMOS database, we defined the best relations to convert COSMOS parameters (coordinates, position angle, diameter, axis ratio and apparent magnitude) into the RC3 system used in the LEDA database. Tiny secondary effects can be tested: distance to plate center effect and air-mass effect.

The converted COSMOS parameters are used to add missing parameters on LEDA galaxies.

## 1 INTRODUCTION

Since 1983 we have been collecting the most important parameters of the brightest galaxies (what we call Principal Galaxies - more than 100,000 such galaxies have already been registered). This is in furtherance of the work of Gerard and Antoinette de Vaucouleurs. To this end, we regularly scan literature for interesting data: accurate coordinates, magnitudes, diameters, position angles, 21-cm line widths, central velocity dispersions, radial velocities, etc. The data are organized in a database called LEDA (Lyon-Meudon Extragalactic Database), which is accessible for free through the international network (via telnet or the electronic mail in batch mode, or via the World-Wide-Web *).

Statistical studies are used to correct the parameters collected to a standard system, basically the system of the Third Reference Catalogue (de Vaucouleurs et al., 1991).

Today, COSMOS data from the "COSMOS/UKST Southern Sky Object Catalogue" (Yentis et al., 1992) are available. They provide many good photometric and astrometric parameters for many galaxies (see H.T. MacGillivray, R.J. Dodd, S.M. Beard, 1987). In order to add these parameters to LEDA galaxies (at least to those which have no equivalent parameters in LEDA), we need to first calculate relations for transforming COSMOS data into RC3 ones. To this end, we extracted a subsample of well measured galaxies from LEDA (i.e. mean error on the apparent total B magnitude lesser or equal to 0.10). This sample contained 1675 galaxies, covering a large range in apparent magnitude and apparent diameters.

The first step in such a comparison consists in cross-identifying COSMOS galaxies with those of our subsample. Thus, in section 2 we explain how this cross-identification was made.

Then, in section 3, we compare parameters provided by COSMOS with their equivalents given by our LEDA subsample. The comparisons are as follows:

• Apparent J magnitudes $B_{cos}$ from COSMOS with total B magnitudes $B_T$ from LEDA

• Apparent diameter $D_{cos}$ from COSMOS with apparent $D_{25}$ diameter at the isophote $25 mag.arcsec^{-2}$ from LEDA. Note that in COSMOS diameters are expressed in arcsec. They are expressed here as log of diameter in 0.1 arcmin, following the definition introduced by de Vaucouleurs, de Vaucouleurs and Corwin (1964).

• Axis ratio $R_{cos}$ (ratio of major axis to minor axis) with $R_{25}$ the axis ratio at the isophote $25 mag.arcsec^{-2}$. As in RC2 and RC3, the comparison is made in log scale.

• Position angle $\beta_{cos}$ from COSMOS with $\beta_{LEDA}$ from LEDA. There is no standard system for the position angle although we may consider that the measurements by Nilson (1973) and Lauberts (1982) for the Northern and Southern hemispheres respectively, constitute a good reference. The position angle is measured in degrees from North Eastwards. Note that the position angle depends on the equinox because it refers to the direction of North. This effect is negligible.

• Equatorial coordinates $\alpha(1950)$, $\delta(1950)$ from COSMOS and LEDA. In LEDA, we have a flag that indicates whether coordinates are accurate or not. In the compari-son we use only accurate coordinates (i.e., coordinates with standard-error lesser than 10 arcsec).

---

\* for internet: telnet lmc.univ-lyon1.fr login: leda
for www http://image.univ-lyon1.fr/base/leda-consult.html

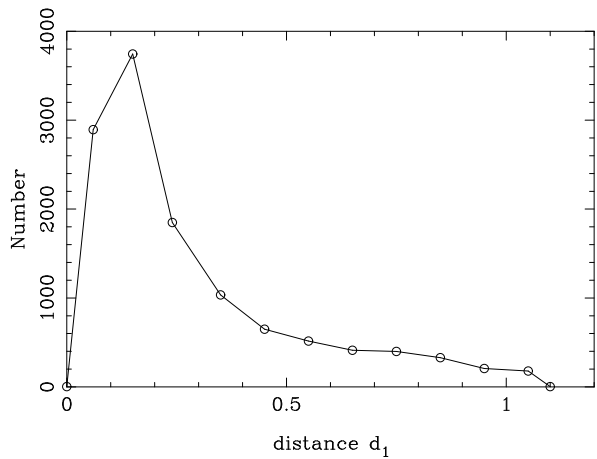

**Figure 1.** Histogram of distances $d_1$ measuring the smallest distance between two objects cross-identified in COSMOS and LEDA

## 2    CROSS-IDENTIFYING COSMOS-LEDA GALAXIES

In recent papers (Paturel et al., 1996; Garnier et al., 1996) we extracted parameters for galaxies identified on 35,000 images digitized from the Palomar Sky Survey. We had to cross-identify LEDA galaxies and galaxies measured on these images. The problem here is similar. So we used the same method. We present the method briefly below (for a more detail presentation, see the papers referred to above).

For each galaxy of the LEDA sample, a distance to COSMOS objects is calculated using not only coordinates but all available parameters with the relation:

$$d_{obj-gal}^{ij} = \frac{1}{N}\sum_{k=1}^{N} w_k . \frac{|p_{ik}-g_{jk}|}{\sigma_i}$$

where N is the number of parameters, $p_{ik}$ is the k-th parameter of the COSMOS object $i$, $g_{jk}$ is the k-th parameter of the LEDA galaxy $j$ (note that these parameters are reduced to the same system following a first comparison).

with the adopted weights :

- $w = 6$    for R.A. and DEC. (or $w = 9$ if accurate)
- $w = 1$    for $logD$, $logR$ and $B_T$
- $w = 2.logR_{cos}$    for position angle $\beta$.

For a given galaxy, the distance of the closest object will be denoted $d_1$ while $d_2 = d_{obj-gal}^{ij}$ will denote the distance for the second closest object ($|d_2 - d_1|$ measures possible confusion). We impose $d_1$ to have a value lesser than 0.6 and the difference $d_2 - d_1$ to have a value larger than a limit (this limit was expressed as a function of $d_1$, typically, $(d_2 - d_1) = 3.d_1 - 0.7$).

We added another condition. For each COSMOS object there is an object-type code: code 1, 5 or 6 for galaxies, code 2 for stars, code 3 or 7 for too faint or multiple objects and code 4 for junk. Only codes 1, 5 or 6 were considered.

## 3    REDUCTION TO THE RC3 STANDARD SYSTEM

The comparison between parameters extracted from COSMOS and standard parameters was made with first order equations.

The LEDA subsample constructed in this way contained 2278 galaxies, 1675 of which were correctly cross-identified with COSMOS galaxies. For the comparison of position angle we added the condition $logR_{25} \geq 0.3$ (for round galaxies the position angle ceased to be meaningful), and for the comparison of coordinates we imposed that they are accurate in LEDA (i.e., with a standard error lesser than 10 arcsec in R.A. or DEC.). In the comparison with the RC3 data we find that the parameters are well represented by linear relationships. The least-square solutions are the following:

$$\beta_{LEDA} = \beta_{cos}$$
$$\sigma = 4.9^{\circ} \qquad n = 394$$

$$(B_T - 13.79) = (B_{cos} - 14.05)$$
$$\sigma = 0.35 mag. \qquad n = 1592$$

$$(logD_{25} - 1.25) = (1.062 \pm 0.007)(logD_{cos} - 0.97)$$
$$\sigma = 0.070 \qquad n = 1361$$

$$logR_{25} = (1.03 \pm 0.01)logR_{cos}$$
$$\sigma = 0.066 \qquad n = 1365$$

In what follows, all COSMOS parameters will be corrected to the RC3 standard system using these relations. Figures 2 and 6 illustrate comparisons of the two systems after correction of COSMOS data.

These results call for some comments. For the position angle we see that a few galaxies are located along the line y=-x. This fact was first noted by Karachensev et al. (1993) and results from some comparison eyeball data measured on badly oriented charts. Such comparison is useful to correct bad position angles collected in LEDA.

We note that the magnitudes for COSMOS data are reasonably reliable over a large range, from B=10 to B=17. If we assume that both magnitude systems have the same mean-error, we infer that the error on $B_{cos}$ is about $0.35/\sqrt{2}$ (i.e., 0.24). However, because we selected LEDA magnitude to have a mean error smaller than 0.10, we conclude that the mean error on COSMOS magnitudes is about 0.33 [†].

The COSMOS diameters correspond to an isophotal level brighter than $25 mag.arcsec^{-2}$. The mean difference $< logD_{25} - logD_{cos} >$ is 0.28. Using the value expressing the change of diameter with surface brightness, $K = \partial logD/\partial \mu = 0.1$ (Fouqué and Paturel, 1985), we conclude that the brightness level of COSMOS diameter is about $22.2 mag.arcsec^{-2}$.

---

[†] $\sigma^2 = \sigma_{cos}^2 + \sigma_{leda}^2$

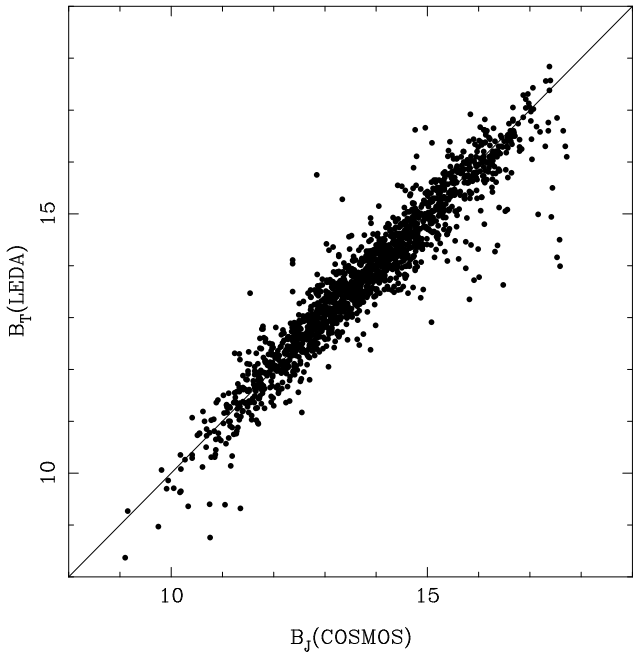

**Figure 2.** Comparison between reduced apparent magnitudes from COSMOS and standard $B_T$ magnitudes.

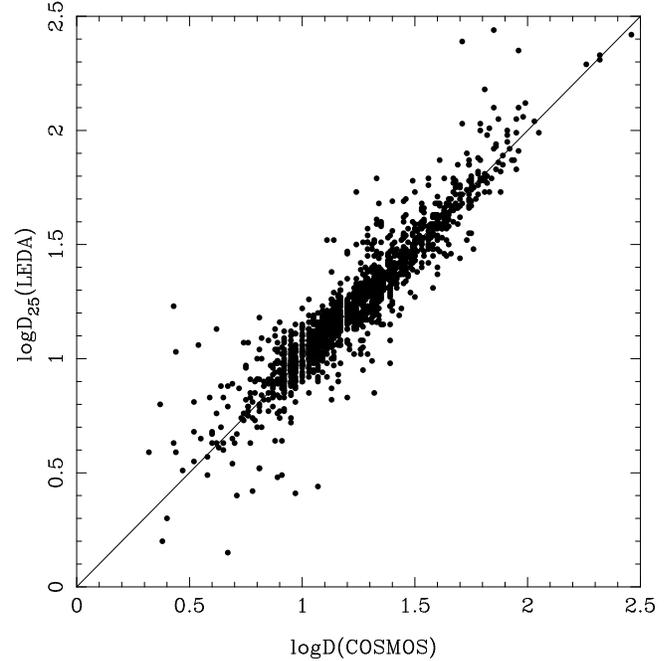

**Figure 3.** Comparison between reduced log of apparent diameters from COSMOS and standard $logD_{25}$.

Concerning coordinates we obtain a mean error of 10.4 arcsec in right ascension and 9.6 arcsec in declination. If that both coordinate systems have the same mean error, we will then conclude that the mean error on position will be about 10 arcsec for COSMOS and LEDA. However, because the mean error on LEDA coordinates can be as large as 10 arcsec according to our selection, one may conclude that COSMOS coordinates are measured with a mean error of a few seconds of arc.

## 4 ADDITIONAL TESTS

We extracted 16973 galaxies having either no total B magnitude or no apparent diameter in LEDA. We made the cross-identification with COSMOS and, finally, obtained a sample of 10543 galaxies, well identified and well documented in COSMOS. This enlarged sample would allow us to search for tiny secondary effects and to test the completeness limit of LEDA after addition of these new data. In our previous paper (Garnier et al., 1996) we found two perturbing effects on magnitudes extracted from PSS-digitized images: a distance to plate center effect and an air-mass effect. We searched for these effects from the 2278 COSMOS-galaxy sample. The first effect was not significant. The second was too large. So, we made the tests again with the whole sample of 12218 COSMOS galaxies (2278+10543).

The test is now significant. The result is quite similar with the one found from PSS images. The least squares solution is (Fig.7):

$(B_T - B_{cos}) = (-0.0004 \pm 0.0001)\Delta r - (0.039 \pm 0.015)$

where $\Delta r$ is the distance (in mm) between the galaxy under consideration and the center of the Schmidt plate (for comparison, the slope found from PSS images was $-0.0006 \pm 0.0002$ and the zero point was $-0.043 \pm 0.032$).

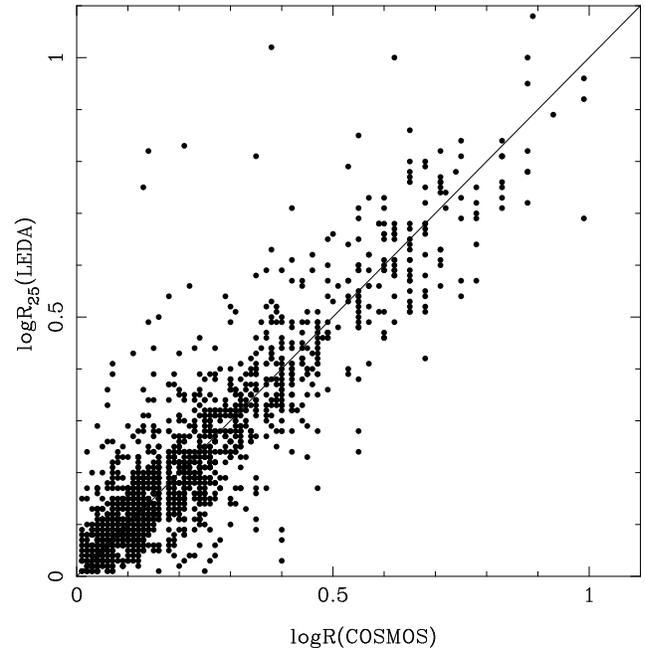

**Figure 4.** Comparison between reduced log of axis ratio from COSMOS and standard $logR_{25}$.

The effect, although present, is quite small, at most no more than 0.09 magnitudes from plate center to plate corner.

Assuming that plates were obtained just at the meridian (sideral time equal to the right ascension), it should be possible to calculate the air-mass of each plate taken at latitude of Siding-Spring Observatory ($-33°$). We can thus search

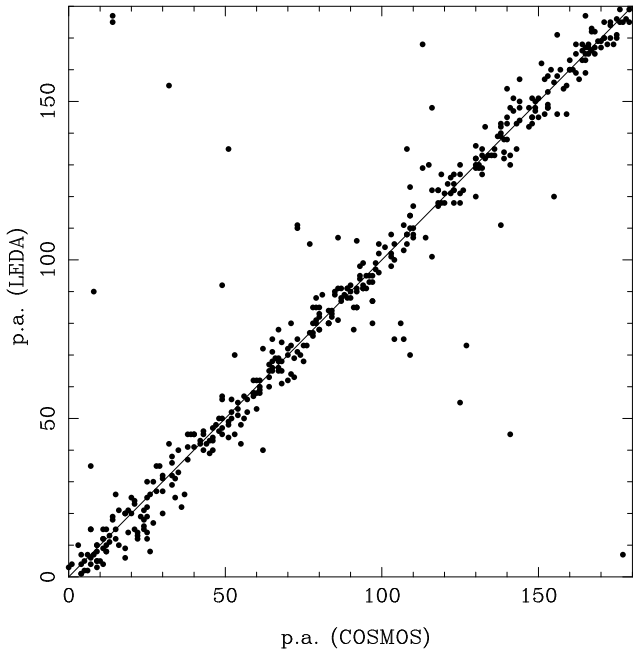

**Figure 5.** Comparison between position angles from COSMOS and from LEDA.

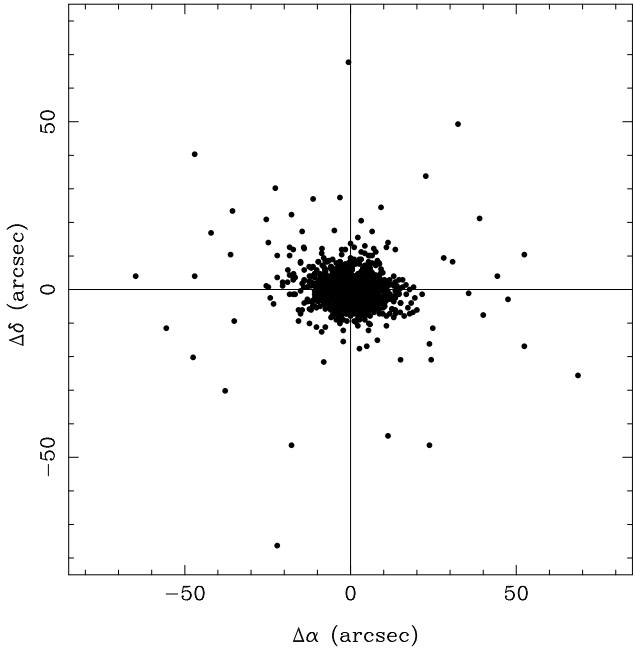

**Figure 6.** Comparison between coordinates from COSMOS and from LEDA.

for an air-mass effect. A least-squares solution leads to a solution:

$$(B_T - B_{cos}) = (-1.37 \pm 0.10) \sec Z + (1.42 \pm 0.11)$$

This solution is quite unrealistic because it assumes an atmospheric extinction of 1.37 in B, whereas a reasonable value is about 0.5. Looking carefully at the distribution of the points we saw that this high value could be explained by some measurements at $\sec Z \approx 1.8$. If these values are rejected, an atmospheric extinction of 0.5 would be acceptable (see Fig.8 on which the dashed line has a slope equal to $-0.5$). This effect would thus be quite similar to that of the PSS in the northern hemisphere (Garnier *et al.* 1996) for which we had a slope of $-0.48 \pm 0.02$. However, one cannot exclude that the effect has another origin (*e.g.*, presence of Magellanic Clouds or obscuration as a function of galactic latitude - both structures becoming more important at these high $\sec Z$ values). Anyway, the effect is quite small and almost negligible over the range $1.00 \leq \sec Z \leq 1.15$.

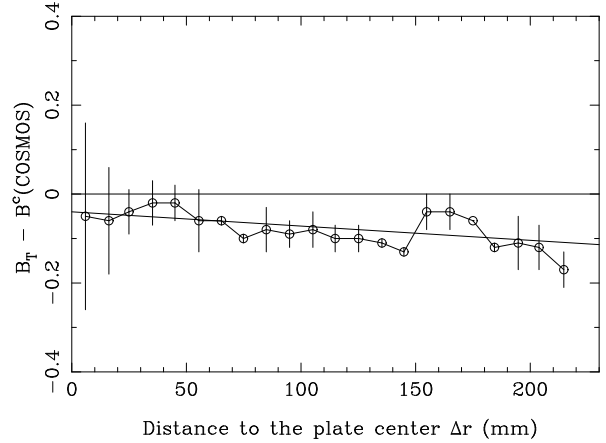

**Figure 7.** Effect on magnitudes of the distance to the center of the chart. This effect is significant although very small. The straight line gives the least-squares solution.

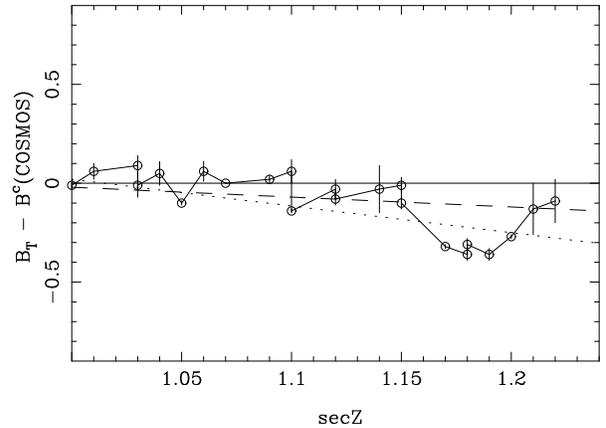

**Figure 8.** Air-mass effect on COSMOS magnitudes. This effect is significant. The doted line gives the raw least-squares solution. The dashed line gives the adopted solution (see text for the discussion).

## 5   COMPLETENESS OF LEDA DATABASE

Tests of completeness of LEDA database are made regularly. These tests are performed by plotting the log of the number of galaxies which are brighter (or larger) than a given apparent magnitude (or diameter) vs. the apparent magnitude (or log of diameter). In a homogeneous universe the expected slope is 0.6 (or $-3.0$ for diameter). In fact the slopes never reach these canonical values (see the discussion in Paturel et al. 1994). These completeness curves are plagued by the fact that many galaxies are identified but have no apparent magnitude and no diameter.

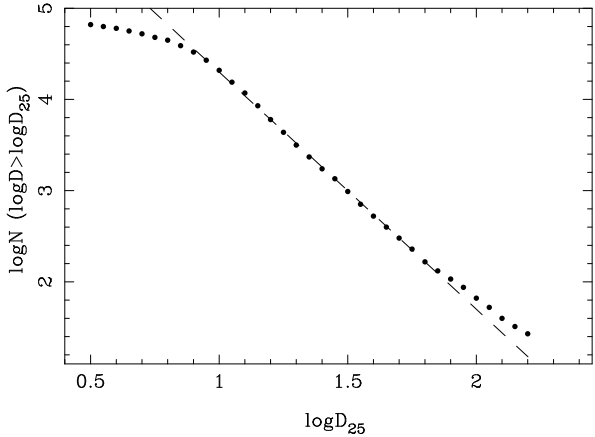

**Figure 9.** Completeness curve in diameter after addition of COS-MOS parameters onto LEDA galaxies. The completeness is full-filled down to $logD_{25} \approx 0.9$.

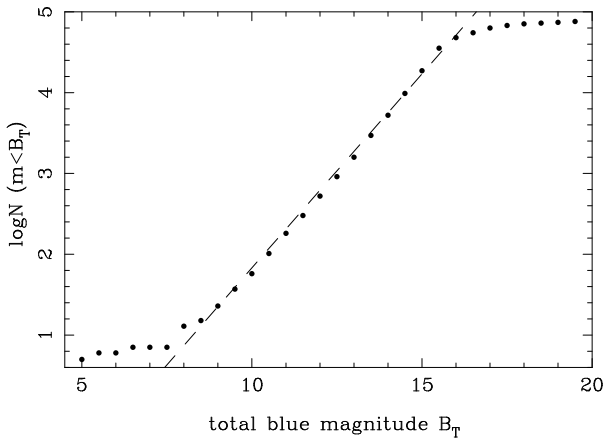

**Figure 10.** Completeness curve in magnitude after addition of COSMOS parameters onto LEDA galaxies. The completeness is fulfilled up to $B_T \approx 16$.

If we include COSMOS data in the Southern hemisphere and of the data we extracted from PSS digitized images in the Northern hemisphere, the completeness of LEDA becomes significantly improved. The completeness curves in magnitude and diameter are presented in Fig.9 and 10. The completeness limit is about 16 in $B_T$ magnitude and 0.9 in $logD_{25}$.

## ACKNOWLEDGEMENTS

We acknowledge the indirect assistance of the COSMOS group at the Royal Observatory who made available the COSMOS/UKST Southern Sky Catalogue used in this study available to the Anglo-Australian Observatory. We want to thank M.C. Marthinet for her technical help and M. Drinkwater for making COSMOS data available at NSW University. We want also want to thank the anonymous referee whose remarks have been very helpful in preparing this final draft.

## REFERENCES

Fouqué P., Paturel G., 1985, A&A **150**, 192
MacGillivray H.T., Dodd R.J., Beard S.M. , 1987, Astronomy from Large Databases, Proceedings eds. F. Murtagh and A. Heck, Garching
Karachentsev I.D., Karachentseva V.E., Parnovsky S.L., 1993, Astron. Nachr., **314**, 97
Lauberts A., 1982, The ESO/Uppsala Survey of the ESO(B) Atlas, European Southern Observatory (ESO)
Nilson P., 1973, Acta Univ. Uppsala, ser. V, vol.1 (UGC)
Paturel G., Bottinelli L., Di Nella H., Fouqué P., Gouguenheim L., Teerikorpi P., 1994, A&A 289, 711
Paturel G., Garnier R., Petit C., Marthinet M.C., 1996, A&A
Garnier R., Paturel G., Petit C., Marthinet M.C., Rousseau, 1996, A&A
Vaucouleurs G. de, Vaucouleurs, A.de, 1964, Bright Galaxy Catalogue, University of Texas Press, Austin
Vaucouleurs G. de, Vaucouleurs, A.de, Corwin, H.G. Jr., Buta, R.J., Paturel, G., Fouqué, P., 1991, Third Reference Catalogue of Bright Galaxies, Springer-Verlag (RC3)
Yentis D.J., Cruddace R.G., Gursky H., Stuart B.V., Wallin J.F., MacGillivray H.T., Collins C.A., 1991, in "Digitalised Optical Sky Surveys", eds. H.T. MacGillivray and E.B. Thomson, Kluwer Academic Publishers, p67